\documentclass{optica-article}

\journal{opticajournal} 

\articletype{Research Article}

\usepackage{xspace}
\usepackage{amsmath}
\usepackage{xcolor}
\usepackage{comment}
\usepackage{multirow}
\usepackage{rotating}
\usepackage{wrapfig}
\usepackage{amsfonts}
\usepackage{graphicx}
\newcommand{\etal}{\textit{et al.\@}\xspace}
\newcommand{\exvivo}{\textit{ex vivo}\xspace}

\newcommand{\invivo}{\textit{in vivo}\xspace}

\newcommand{\invitro}{\textit{in vitro}\xspace}

\newcommand{\enface}{\textit{en face}\xspace}

\newcommand{\um}{\(\muup\)m\xspace}

\newcommand{\rr}{{\boldsymbol{r}\xspace}}
\newcommand{\rhorho}{\boldsymbol{\rho}\xspace}

\newcommand{\sillps}{{\mathrm{S_{ill}^{ps}}\xspace}}
\newcommand{\scolps}{{\mathrm{S_{col}^{ps}}\xspace}}
\newcommand{\pillps}{{\mathrm{P_{ill}^{ps}}\xspace}}
\newcommand{\pcolps}{{\mathrm{P_{col}^{ps}}\xspace}}
\newcommand{\sillsc}{{\mathrm{S_{ill}^{sc}}\xspace}}
\newcommand{\scolsc}{{\mathrm{S_{col}^{sc}}\xspace}}
\newcommand{\pillsc}{{\mathrm{P_{ill}^{sc}}\xspace}}
\newcommand{\pcolsc}{{\mathrm{P_{col}^{sc}}\xspace}}
\newcommand{\psfps}{\mathrm{PSF}_{\mathrm{ps}}\xspace}
\newcommand{\psfsc}{\mathrm{PSF}_{\mathrm{sc}}\xspace}

\newcommand{\amp}{\mathrm{\alpha}}
\newcommand{\ampps}{\mathrm{\alpha}_{\mathrm{ps}}}
\newcommand{\ampsc}{\mathrm{\alpha}_{\mathrm{sc}}}
\newcommand{\ips}{{\mathrm{I}_\mathrm{ps}\xspace}}

\newcommand{\apps}{\mathrm{A}_p^\mathrm{ps}\xspace}
\newcommand{\apsc}{\mathrm{A}_p^\mathrm{sc}\xspace}

\newcommand{\phaseps}{{\varphi_{\mathrm{ps}}}\xspace}
\newcommand{\phasesc}{{\varphi_{\mathrm{sc}}}\xspace}

\newcommand{\scFFOCT}{{spatially-coherent FFOCT}\xspace}
\newcommand{\ScFFOCT}{{Spatially-coherent FFOCT}\xspace}

\graphicspath{{./figures/}}
\DeclareMathOperator{\sgn}{sgn}

\begin{document}

\title{Theoretical analysis of performance limitation of computational refocusing in optical coherence tomography}

\author{Yue Zhu\authormark{1},
 		Shuichi Makita\authormark{1},
 		Naoki Fukutake\authormark{2},
		and Yoshiaki Yasuno\authormark{1,*}}

\address{
\authormark{2}University of Tsukuba, Tennodai 1–1-1, Tsukuba, Ibaraki, 305-8573, Japan.\\
	\authormark{3}Nikon Corporation, 471 Nagaodai-cho, Sakae-ku, Yokohama-city, Kanagawa, 244-8533, Japan.
}

\email{\authormark{*}yoshiaki.yasuno@cog-labs.org} 

\begin{abstract*}
High-numerical-aperture optical coherence tomography (OCT) enables sub-cellular imaging but faces a trade-off between lateral resolution and depth of focus. 
Computational refocusing can correct defocus in Fourier-domain OCT, yet its limitations remain unaddressed theoretically. 
We formulate the lateral imaging process of OCT by using pupil-based imaging theory and the constraints of the computational refocusing in point-scanning OCT and spatially-coherent full-field OCT (FFOCT) are analyzed.
The constrains in lateral sampling density and the confocality are considered, and it is shown that the maximum correctable defocus (MCD) is primarily limited by confocality in point-scanning OCT, while \scFFOCT has no such constraint and can achieve virtually infinite MCD with a proper and reasonable sampling density.
This makes \scFFOCT particularly suitable for optical coherence microscopy.
\end{abstract*}

\section{Introduction}
As a result of the progress in cultivation technology, \invitro samples have come to emulate living tissues more realistically, and thus these samples are becoming thicker. 
These samples and also thick \exvivo and dissected tissues have been widely used for biology and medicine, and simultaneous cellular-level-resolution and millimeter-depth imaging is essential to investigate these samples.
The optical coherence tomography (OCT) microscopy\cite{Izatt1994OL, Beaurepaire1998OL, Aguirre2003OL} and its variants such as dynamic OCT microscope \cite{Apelian2016BOE, Muenter2020OL, Leung2020BOE, ElSadek2020BOE} have been demonstrated for non-invasive high-resolution imaging of \invitro samples, such as spheroids \cite{Huang2017CancRes, ElSadek2020BOE, ElSadek2021BOE, ElSadek2024SciRep, KYChen2024BOE} and organoids \cite{Scholler2020LSA, Morishita2023BOE}, dissected tissues \cite{Izatt1996JSTQE, YuChen2007JBO, Muenter2020OL, Leung2020BOE, Mukherjee2022BOE, Pradipta2023SciRep, Zhao2023ACSNano}, and also \invivo tissues \cite{Bizheva2017BOE, Bizheva2024TVST}.

However, similar to other optical microscopic modalities, OCT suffers from a trade-off between its lateral resolution and its depth-of-focus (DOF).
Namely, a high-numerical-aperture (high-NA) objective can provide high lateral resolution, but its DOF becomes shallow.
As the target samples became thicker, this resolution-DOF trade-off became more the important problem of OCT microscopes.

A short DOF is not a major issue for time-domain OCT microscope, because it measures each depth at each image acquisition.
Specifically, we can introduce a technique known as ``dynamic focus''\cite{Lexer1999JMO, Michael2006JBO, Dubois2006OC}, which shifts the depth position of the focus to the optimal depth for acquisition of each individual image.
On the other hand, Fourier-domain (FD-) OCT with a high NA objective, i.e., OCT microscope, is not free from the resolution-DOF trade-off.

It is known that this trade-off can be mitigated using computational refocusing methods, such as holographic focus and/or aberration correction \cite{Yasuno2006OE, Kumar2013OpEx, Kumar2021BOE} or interferometric synthetic aperture microscope (ISAM) \cite{Raston2007NatPhy} techniques.
The former method manipulates the phases of the spatial frequency spectrum of complex OCT signals to remove the defocus.
The latter method (ISAM) first converts a complex OCT signal into its spatial frequency spectrum and then resamples the spectrum into an appropriate space to remove the defocus.
A combination of the holographic signal processing and ISAM has also been demonstrated \cite{Adie2012NatAcaSci}.

Despite the increasing importance of OCT microscopy and the computational refocusing, these computational refocusing methods are expected to have certain limitations.
In holographic refocusing, the spatial sampling density of the OCT signal limits the accessible spatial frequency component and thus may govern the maximum correctable defocus.
Specifically, larger defocus causes larger local phase-error slope, and hence, consists of higher spatial frequency information.
To capture this higher spatial frequency information, higher spatial sampling density are required.
The confocality may also limit the practical correctable defocus because it causes an optical energy loss, and thus, a drop in the signal-to-noise ratio (SNR), at depths far from the depth position of the physical focus.
Computational refocusing techniques, including both holographic refocusing and ISAM, cannot recover this optical energy loss.
Although these limitations have been anticipated, they have not been theorized well or investigated thoroughly.
In addition, the effects of these limiting factors and the associated affecting mechanisms vary among the different types of OCT systems, e.g., scanning OCT and full-field OCT.

This work aims to establish a theoretical framework that can predict the maximum correctable defocus for multiple types of OCT, including conventional point-scanning OCT and full-field (FF-) swept-source OCT with spatially coherent illumination (known hereafter as spatially-coherent FFOCT).
We start with mathematical descriptions of the image formation processes for these types of OCT.
Two standard OCT types, i. e., point-scanning OCT and spatially-coherent FFOCT, are considered. 
Here, we use the dual pupil-based formulation for OCT imaging \cite{Ralston2005IEEE}.
Then, the defocus in the two OCT types under study is described mathematically by extending this formulation.
Finally, the criteria for the maximum correctable defocus are derived for both the point-scanning OCT and \scFFOCT.
In this work, we consider two limiting factor types.
The first is the sampling-density limit.
Because it is related to the Nyquist frequency required to detect and correct the phase errors induced by the defocus properly, we call it the ``Nyquist criterion.''
The other factor is the confocality limit.
Because it is related to the signal reduction at the out-of-focus depth, we call this factor the ``confocality-limit criterion.''
Our theory will indicate that \scFFOCT has a virtually infinitely large maximum correctable defocus and is thus particularly suitable for OCT microscopy applications.

\section{Theoretical framework for comprehension of OCT image formation}
\subsection{Two-dimensional pupil-based imaging formation theory}
To derive the phase sampling limit, we start from mathematical modeling of the two-dimensional (2D) lateral imaging process of OCT\@.
Here the imaging process is described using the concepts of the ``conceptual pupil'', the ``spot'',the ``aperture'', and the ``point spread function (PSF)''\cite{Fukutake2020SciRep,fukutake_four-dimensional_2025-1,fukutake_unified_2025-1}.
For simplicity, we assume that the lateral and axial resolutions are not coupled, thus allowing us to focus our consideration on the \enface plane.
Note that the 2D nature of this analysis limits its applicability to the \enface signal processing-based holographic refocusing approach, and it is not applicable to ISAM because ISAM is based on 3D remapping of data in the spatial frequency domain.

\begin{figure}
	\begin{center}
		\includegraphics[width=0.8 \textwidth]{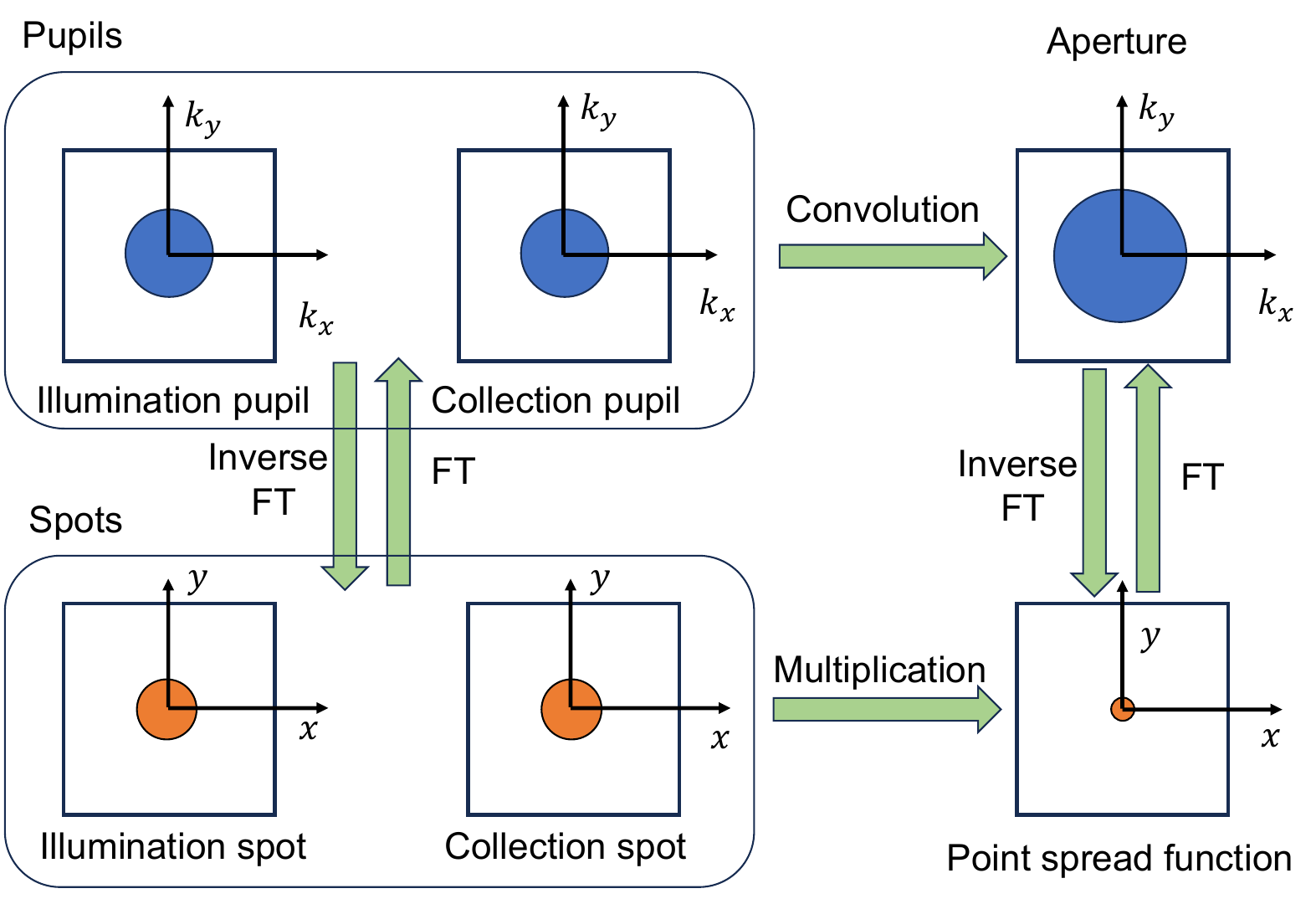}
	\end{center}
	\caption{
		Diagram illustrating the interrelationship among the conceptual pupils, spots, aperture, and point spread function.
		$(x,y)$ and $(k_x, k_y)$ are the lateral spatial coordinates and their corresponding spatial frequencies, respectively.
		FT stands for Fourier transform.
} 
	\label{fig:probe_optics}
\end{figure}

Because we are modeling the 2D lateral imaging process, a 2D (i.e., not 3D) conceptual pupil is used.
Figure \ref{fig:probe_optics} presents a diagram that illustrates the conceptual pupil theory. 

Although the conceptual pupil is somehow related to a physical pupil in an optical system, it does not really correspond to the physically existing pupil or to a particular plane (i.e., the pupil plane) in the optics.
The conceptual pupil is a representation of illumination or collection (i.e., detection) optics in the spatial frequency domain.
The extension (i.e., the size) of the conceptual pupil in the frequency domain is defined by the NA of the optics, where a larger the NA corresponds to a larger the conceptual pupil.
In addition, the wavefront aberration, which includes the defocus, is represented as the phase of the conceptual pupil.
Because each of illumination and collection optics has its own conceptual pupil, there are two conceptual pupils in a single OCT system, i.e., the conceptual illumination and collection pupils, as illustrated at the top left of Fig.\@ \ref{fig:probe_optics}.
Hereafter, we denote the conceptual pupil as ``pupil'' for simplicity.

The ``spot'' is defined as the 2D inverse Fourier transform (inverse FT) of the pupil.
The inverse FT of the illumination pupil is called the illumination spot and is the real beam spot on the sample.
On the other hand, the inverse FT of the collection pupil gives us a virtual spot on the sample, which is referred to as the ``collection spot.''
The collection spot is an spot that might be observed on the sample if we inverted the light propagation direction, i.e., from the detector toward the lightsource.
It can be readily imagined that if wavefront aberrations such as defocus occur in the optics, they will cause widening of the spot because the aberration appears as the phase error of the pupil.

The ``aperture'' is defined as the convolution of the illumination and collection pupils in the spatial frequency domain, and it corresponds to the total imaging system.
The aperture restrains and modulates the spatial frequency spectrum of the sample and only the spectrum that has been modified by the aperture is detected.
Namely, the spatial frequency spectrum of the sample is multiplied by the aperture in the spatial frequency domain before the photodetection process.

The PSF can be derived in two ways.
In the first way, the PSF is derived (or defined) as the inverse FT of the aperture.
Alternatively, the PSF can be derived (or defined) as the product of the illumination and collection spots.
Both derivations are mathematically equivalent because of the convolution theory of the FT.
It is well known that PSF defines the resolution of OCT\cite{Drexler2015OCT}.

\subsection{Point-scanning OCT and \scFFOCT}
\label{sec:psfOfOct}
In this manuscript, two types of OCT are considered: fiber-based point-scanning OCT and \scFFOCT.
\ScFFOCT is typically implemented as a swept-source OCT with plane-wave illumination.
Note that time-domain FFOCT typically uses a spatially incoherent light source, and thus the discussion in this manuscript cannot be applicable to the spatially-incoherent time-domain FFOCT.
On the other hand, the point-scanning OCT system in this manuscript covers time-domain, spectral-domain, and swept-source OCT systems as far as they are single-mode-fiber-based point-scanning systems.

\begin{figure}
	\begin{center}
		\includegraphics[width=1.0 \textwidth]{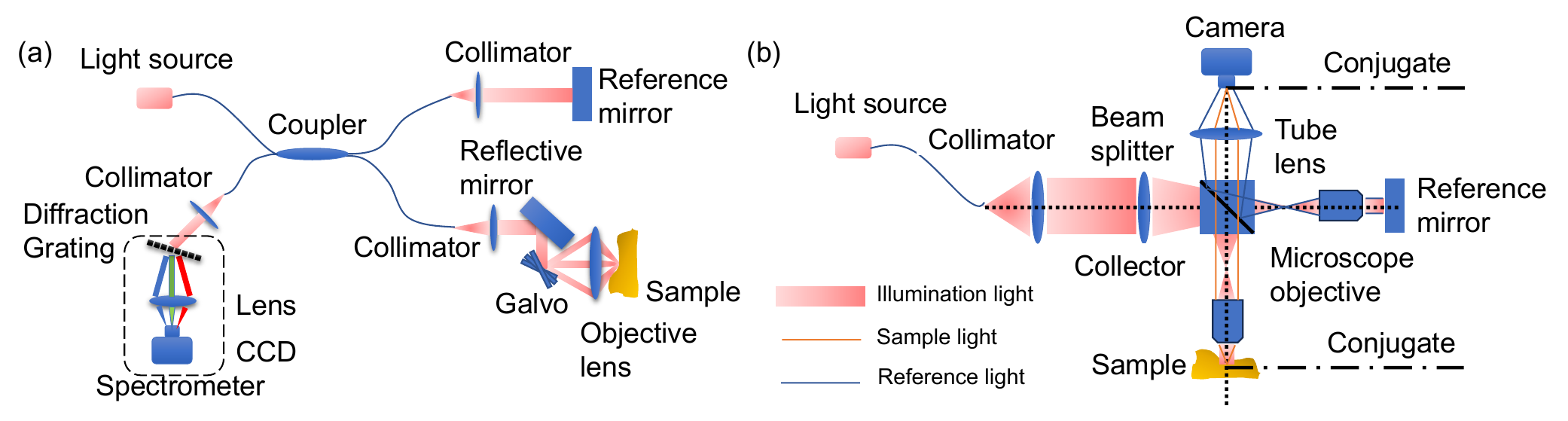}
	\end{center}
	\caption{Configurations of (a) point-scanning OCT and (b) spatially-coherent FFOCT.}
	\label{fig:Systems}
\end{figure}

Figure \ref{fig:Systems} shows configurations of the two OCT system types.
The point-scanning OCT forms a single probe beam spot on the sample and performs three-dimensional tomography of a sample via a two-dimensional lateral scanning, which is typically performed by a galvanometer scanner.
On the other hand, the \scFFOCT illuminates the sample with collimated light (i.e., a plane wave) and then images the sample using a two-dimensional camera. 
This means that the sample plane and camera plane are optically conjugated.
If \scFFOCT was implemented as a swept-source OCT system, each \enface point on the sample, and similarly each pixel of the camera, would be resolved for the wavelength, and the depth resolution was obtained via the standard tomographic reconstruction method used in Fourier-domain OCT \cite{Leitgeb2004OE}.

We can reasonably assume that the reference beam of the OCT setup does not have a lateral structure, i.e., it is constant over the lateral imaging field.
Specifically, the reference light field for \scFFOCT is a plane wave on the 2D camera.
Based on this assumption, the lateral imaging properties of OCT are fully governed by the illumination and collection optics.
In other words, these properties are fully governed by the illumination and collection pupils.

\subsubsection{Point-scanning OCT}
	\label{sec:psfps}
The illumination optics of point-scanning OCT start from the single-mode fiber tip. 
A diffraction-limit Gaussian beam is incident from the fiber tip, diverges as it propagates, and is then collimated by a collimator to form a collimated Gaussian beam. 
This Gaussian beam illuminates the objective and then forms a focused beam spot on a sample.
Because the physical aperture of the objective is larger than the collimated Gaussian beam, the illumination spot becomes Gaussian.
This beam spot is identical to the illumination spot $\sillps(\rr;k)$, where $\rr=(x,y)$ represents the lateral coordination in the real (i.e., physical) space and $k = 2\pi/\lambda$ is the wavenumber corresponds to the representative wavelength $\lambda$ of the illumination. Hereafter, we omit $k$ for simplicity, and this omission does not cause significant inaccuracy under relatively narrow-band approximation.

The illumination pupil $\pillps(\rhorho)$ is a Fourier transform of the illumination spot, where $\rhorho=(k_x,k_y)$ represents the lateral spatial frequency.
In other words, the illumination spot is the inverse Fourier transform of the illumination pupil. 
As a result, a larger illumination pupil corresponds to a smaller illumination spot.
In addition, because the spot is Gaussian, the pupil will also have a Gaussian shape.
(Note that the Fourier transform of a Gaussian function is also a Gaussian function.)

Because the point-scanning OCT systems shares the same optics for illumination and collection (i.e. light detection), the collection pupil $\pcolps(\rhorho)$ and spot $\scolps(\rr)$ become identical to those of the illumination process. 

The PSF is the product of the illumination and collection spots, the PSF of the point-scanning OCT system becomes
\begin{equation}
	\label{eq:psfPs}
	\psfps(\rr;k) = \sillps(\rr;k) \scolps(\rr;k) = G^2(\rr;k),
\end{equation}	
where $G(\rr;k) = \sillps(\rr;k) = \scolps(\rr;k)$ is a Gaussian function. 
Because the squared Gaussian function also becomes a Gaussian, the $\psfps$ is a Gaussian, and it is $\sqrt{2}$-times sharper than the illumination and collection spots in terms of both their amplitudes and their squared intensities.

The aperture of the point-scanning OCT can be derived either by convolving the two pupils as $\apps = \pillps (\rhorho; k)* \pcolps (\rhorho; k) = G(\rhorho; k) *  G(\rhorho; k)$ or by Fourier transforming the PSF as $\psfps(\rr;k)$.
Because the convolution of two Gaussian functions produces a Gaussian function and the Fourier transform of a Gaussian function is also a Gaussian function, the aperture of the point-scanning OCT system is also a Gaussian.
Note that the Gaussian aperture extends to an infinity high spatial frequency without any apparent cut-off frequency in this model.
This scenario corresponds to our assumption that the physical aperture is sufficiently larger than the collimated Gaussian beam. 
This assumption is reasonable for most current point-scanning OCT systems.

\subsubsection{\ScFFOCT}
\label{sec:psfsc}
While conventional time-domain FF-OCT uses incoherent flood illumination, \scFFOCT illuminates the sample using a spatially coherent plane wave, i.e., a light field with a flat phase.
This indicates that the light source should be fully spatially coherent, and this condition can be achieved when the light is incident from a single-mode fiber tip, as shown in Fig.\@ \ref{fig:Systems}(b).
The light is then collimated once and converged at the back focal plane of the objective, which means that it is collimated again by the objective and thus illuminates the sample as a plane wave.
Specifically, the illumination spot of the \scFFOCT is a constant as $\sillsc(\rr;k) = \mathrm{Constant}$.

Because the illumination spot is a constant, the illumination pupil, which is given by the Fourier transform of the spot, then becomes a delta function
$\pillsc = \delta(\rhorho; k)$.
 
In practical \scFFOCT systems, the objective has a physical aperture with a specific size.
This limits the collectable spatial frequency, and as a result, the collection pupil becomes a cylinder function with a specific cut-off frequency.
This cut-off frequency is governed by the NA of the objective, where a larger NA results in a higher cut-off frequency.

The inverse Fourier transform of the cylinder function is an Airy disk function, and thus the in-focus collection spot of the \scFFOCT is also represented by an Airy disk function.  Here, we ignore the relatively small outer rings of the Airy disk and can approximate the central lobe reasonably well using a Gaussian profile \cite{Zhang2007AO}.
The out-of-focus collection spot is assumed to be the convolution of the approximated Gaussian in-focus spot and the phase function.
Namely, we consider a virtual Gaussian collection spot $\scolsc(\rr;k)$ for the \scFFOCT, which is similar to the collection spot used for point-scanning OCT\@.
It should be noted here that this Gaussian approximation results in a tacit approximation that the collection pupil $\scolsc(\rr;k)$ also has a Gaussian.

Because the illumination spot for the \scFFOCT is a constant, the PSF, which is given by the product of the illumination and collection spots, becomes identical to the collection spot because
\begin{equation}
\psfsc(\rr; k) = \scolsc(\rr; k).
\end{equation}
Similarly, because the illumination pupil is a delta function, the aperture for \scFFOCT, which is given by the convolution of the illumination and collection pupils, becomes identical to collection pupil as $\apsc(\rhorho; k) = \pcolsc(\rhorho; k)$.
The aperture can also be considered to be the Fourier transform of the PSF.
Because the PSF is identical to the collection spot, the same conclusion can be derived from this definition.

\subsection{Defocus in point-scanning OCT and \scFFOCT}
The pupil-based theoretical modeling approach in Section \ref{sec:psfOfOct} clarified the PSFs for both point-scanning OCT and \scFFOCT, and also clarified their relations.
We now can describe the differences between defocus effects in point-scanning and \scFFOCT.
\subsubsection{Defocused Gaussian beam in point-scanning OCT}
\label{sec:PSF_defocus}
\begin{figure}
	\begin{center}
		\includegraphics[width=1 \textwidth]{Optics.pdf}
	\end{center}
	\caption{Schematic diagrams of probe optics used in (a) point-scanning OCT and (b) \scFFOCT.}
	\label{fig:Optics}
\end{figure}	
Figure \ref{fig:Optics}(a) shows a schematic diagram of probe optics used in point-scanning OCT.
In this setup, the illumination and collection path share the same optics, and thus the illumination and collection spots are identical. 
In addition, because the probe beam emerged from a single-mode fiber, the spots have Gaussian profiles.

The sample plane (i.e., the \enface imaging plane) is assumed to be shifted from the focus depth by $z_d$.
Hereafter we call $z_d$ as the ``defocus distance.''
The Gaussian spot, which represents the illumination and collection spots equally, can be given by using the Gaussian beam \cite{kogelnik_laser_1966} as
\begin{equation}
	\label{eq:G_col}
	G(\rr;z_d) = \frac{w_0}{w(z_d)} \exp \left[-\frac{\rr^2}{w^2(z_d)}\right]
	\exp
	\left[
	-i\left\{
	nk_0(z_d)+\frac{nk_0\rr\cdot\rr}{2R(z_d)}-\psi(z_d)
	\right\} 
	\right],
\end{equation}
where $w_0$ is the beam waist radius of the amplitude (not the intensity) at the in-focus depth, i. e., $2w_0$ is the diffraction-limit $1/e$-width spot size of the amplitude, and equally, the $1/e^2$-width spot size of the squared intensity of the spot.
More specifically, $w_0 = 4f/\phi k_0$, where $k_0 = \frac{2\pi}{\lambda_0}$ is the wave number that corresponds to the center wavelength $\lambda_0$, $n$ is the refractive index of the sample, $f$ is the focal length of the objective and $\phi$ is the $1/e^2$-diameter of the probe beam incident at the objective.
$w(z_d)$ is the beam radius with a particular defocus $z_d$, and is given by
\begin{equation}
\label{eq:beamradius}
	w(z_d) = w_{0}\sqrt{1+{z_d}^2/{z_R}^2},
\end{equation}
where $z_R=\left(n k_0 {w_0}^2\right)/2=\left(n {w_0}^2 \pi\right)/\lambda_0$ is the Rayleigh length.
The second term in the second exponential represents the quadratic phase induced by defocus and $R(z_d)$ is the phase curvature:
\begin{equation}
	\label{eq:phaseCurvature}
	R(z_d) = z_d\left[1+{z_R^2}/{z_d^2}\right].
\end{equation}
In the third term of the second exponential, $\psi(z_d)$ is the Gouy phase.

As discussed in Section \ref{sec:psfOfOct} the PSF is the product of the illumination and collection spots, i.e., $G^2(\rr;z_d)$ [Eq.\@ (\ref{eq:psfPs})].
As a result, the amplitude profile $\amp(\rr; z_d)$ and the $\rr$-dependent phase term induced by the defocus $\phaseps$ become
\begin{equation}
	\label{eq:psfAmpPs}
	\amp(\rr; z_d) \propto
		\exp \left[
			-\frac{2 \rr\cdot\rr }{w^2(z_d)}
		\right],
\end{equation}
and
\begin{equation}
	\label{eq:psfPhasePs}
	\phaseps(\rr; z_d) = \frac{nk_0\rr \cdot \rr}{R(z_d)},
\end{equation}
as
\begin{equation}
	\label{eq:psfPsBroken}
	\psfps(\rr;z_d) \propto \ampps(\rr; z_d) \exp\left[i \phaseps (\rr; z_d)\right].
\end{equation}

\subsubsection{Full-field collected defocus in SC-FFOCT}
Figure \ref{fig:Optics}(b) illustrates the illumination and collection used for \scFFOCT.
Because the illumination is a plane wave, the illumination spot remains a constant, irrespective of the defocus distance $z_d$ (see also Section \ref{sec:psfsc}.)
Note that this insensitivity of the illumination spot to the defocus distance can be also described with respect to the illumination pupil.
In general, the defocus can be described as the phase error of the pupil.
Because the illumination pupil for \scFFOCT ($\pillsc$) is a delta function, any phase error only causes an constant phase offset.
Because the illumination spot is the inverse Fourier transform of the illumination pupil, the illumination pupil is thus not sensitive to the defocus, with the exception of a possible constant phase offset.

The collection spot is approximated as a Gaussian spot in our model (see Section \ref{sec:psfsc}), and we can use the collection spot for the point-scanning OCT [Eq.\@ (\ref{eq:G_col})].
As a result, the PSF of the \scFFOCT, which is the product of the illumination and collection spots, becomes
\begin{equation}
	\psfsc(\rr;z_d) = \mathrm{C} \, G(\rr;z_d) \propto G(\rr;z_d),
\end{equation}
where $\mathrm{C}$ is a complex constant that represents the illumination spot.

Therefore, the amplitude profile $\ampsc(\rr; z_d)$ and the $\rr$-dependent phase term induced by the defocus $\phasesc$ become
\begin{equation}
	\label{eq:psfAmpSc}
	\ampsc(\rr; z_d) \propto \exp \left[-\frac{\rr\cdot\rr}{w^2(z_d)}\right],
\end{equation}
and
\begin{equation}
	\label{eq:psfPhaseSc}
	\phasesc (\rr; z_d) = \frac{nk_0\rr\cdot\rr}{2R(z_d)},
\end{equation}
as
\begin{equation}
	\label{eq:psfScBroken}
	\psfsc(\rr;z_d) \propto \ampsc(\rr; z_d) \exp\left[i \phasesc (\rr; z_d)\right].
\end{equation}

With the normalized defocus distance $\zeta_d = z_d/z_R$, substitution of $R(z_d)$ [Eq.\@ (\ref{eq:phaseCurvature})] into $\phaseps$ and $\phasesc$ means that these phase, as functions of the normalzied defocus distance, then becomes,
$\phaseps(\rr; \zeta_d) = \frac{2\rr \cdot \rr}{w_0^2}\frac{1}{\zeta_d [1+1/\zeta_d^2]}$ 
and
$\phasesc(\rr; \zeta_d) = \frac{\rr \cdot \rr}{w_0^2}\frac{1}{\zeta_d [1+1/\zeta_d^2]}$, respectively.
Comparison of these equations with the same defocus distance shows that the defocus-induced phase of \scFFOCT is two times smaller than that of point-scanning OCT. Therefore, it can be deducted that the $\rr$-dependent phase of the PSF for \scFFOCT is two times less sensitive to the defocus distance than that of the PSF for point-scanning OCT.

\section{Criteria for maximum correctable defocus}
Two factors limit the maximum correctable amount of defocus: the lateral image sampling density and the confocality.
The former factor both the point-scanning OCT and \scFFOCT, whereas the latter factor only affects point-scanning OCT.
The maximum correctable defocus amounts defined by these factors are derived in the following sections.

\subsection{Lateral sampling density limit for maximum correctable defocus}
\label{sec:NyquistCriteria}
To correct the defocus via holographic refocusing, the complex OCT data should be sampled with a sufficiently high lateral data density, i.e., the lateral spatial sampling frequency should be higher than the maximum spatial frequency spectrum of the PSF.
This ``Nyquist criterion'' is the necessary and sufficient criterion for the lateral sampling density.
For ease of understanding of the derivation, we first derive the Nyquist criterion for point-scanning OCT and then derive corresponding criterion for the \scFFOCT.

\subsubsection{Nyquist criterion for point-scanning OCT}
\label{sec:NyquistPs}
The PSF of point-scanning OCT [Eq.\@ (\ref{eq:psfPsBroken})] is given by the product of the real Gaussian amplitude [Eq.\@ (\ref{eq:psfAmpPs})] and the phase-only function $\exp\left[i \phaseps(\rr; z_d)\right]$, where $\phaseps$ is the phase defined in Eq.\@ (\ref{eq:psfPhasePs}).
Therefore, the spatial frequency spectrum of the PSF is given by convolution of the Fourier transform of the real Gaussian amplitude and the Fourier transform of the phase-only function.
As the defocus increases, the real Gaussian amplitude becomes broader, and thus, its spatial frequency spectrum becomes narrower.
On the other hand, as the defocus increases, the phase-only function consists of higher frequency components,  especially at the periphery (i.e., at larger $r$, where $r = \left|\rr \right|= \sqrt{x^2 + y^2}$).
As a result, the Nyquist criterion is governed by the Nyquist frequency of the phase-only function in this case.

The phase-only function is a quadratic function of $\rr$, and its local frequency increases as $r$ increases.
To sample the OCT signal to allow it to be refocused, the lateral sampling density should be high enough when compared with the local frequency.
For the phase-only function, the Nyquist condition can be described as follows: ``the adjacent sampling points should have a phase difference smaller than or equal to $\pi$,'' which can be written as
\begin{equation}
	\label{eq:ParDerivative}
	\left|\Delta \phaseps(x;z_d)\right|
		= \left| \frac{\partial}{\partial x}\left( \frac{nk_0 x^2}{R(z_d)} \right) \Delta x\right|
		= \left|\frac{2nk_0 x}{R(z_d)} \Delta x\right| \le\pi,
\end{equation}
where we replaced $\rr$ with $x$ without losing generality. Here, $\Delta \phaseps$ is the phase difference between adjacent sampling points around $x$ and $\Delta x$ is the lateral sampling distance.
$x$ is a generalized lateral position which can be in any lateral direction.
We also assume $y$ is the counter part of $x$ and is oriented along the direction orthogonal to $x$.
The origin of the $(x,y)$ coordinates is collocated with the center of the PSF.

As Eq.\@ (\ref{eq:ParDerivative}) shows, the absolute phase increments linearly increases by $x$, and thus it reaches a maximum at the periphery of the PSF. 
Here, we can reasonably define the radius of the PSF as the $1/e$-radius of the amplitude, i. e., $w(z_d)/\sqrt{2}$, and thus, the maximum absolute phase increment, which is observed at the periphery of the PSF, becomes
\begin{equation}
	\label{eq:PS_max}
	\mathrm{max}\left(\left|\Delta \phaseps (z_d; x)\right|\right)_x
	= \left| \frac{\sqrt{2}nk_0 w(z_d)}{R(z_d)}\Delta x\right|,
\end{equation}
where $\mathrm{max}(\quad)_x$ represents the maximum over $x$, and this maximum is obtained at $x = w(z_d)/\sqrt{2}$.
Now we consider the $x$ as a parameter with a certain value, while $z_d$ will be treated as a variable rather than a parameter.

\begin{figure}
	\begin{center}
		\includegraphics[width=0.65 \textwidth]{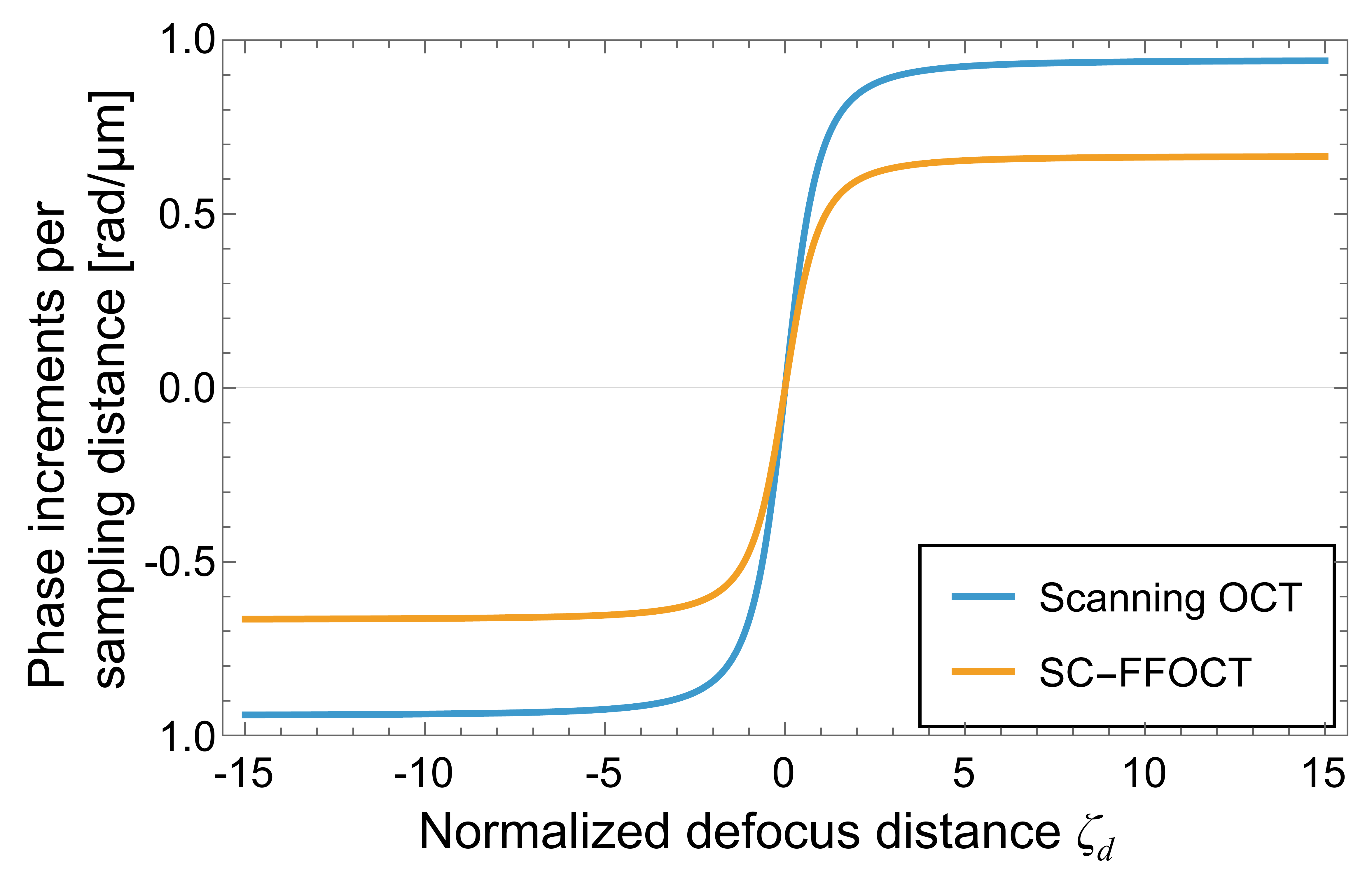}
	\end{center}
	\caption{The phase increments per the sampling distance at the periphery of the PSF, which corresponds to Eq.\@ (\ref{eq:MaxPsXi}) and Eq.\@ (\ref{eq:MaxScXi}) but without absolute operations.
		The blue and orange curves correspond to the cases of point-scanning OCT and \scFFOCT, respectively.
		The horizontal axis corresponds to the normalized defocus distance $\zeta_d$, which takes a value of $\pm 1$ when the absolute defocus distance is equal to the Rayleigh distance.
		The phase increments takes its maximum and minimum values when $\zeta_d$ approaches to $+\infty$ and $-\infty$, respectively.
		For the figure plot, we assumed $w_0$ = 3 \um.
}
	\label{fig:piMaxPhaseCurvDependence}
\end{figure}
This equation can be rewritten using the normalized defocus distance $\zeta_d= {z_d} / {z_R}$, which is the defocus distance normalized with respect to the Rayleigh length, as
\begin{equation}
	\label{eq:MaxPsXi}
	\mathrm{max}\left(\left|\Delta \phaseps (\zeta_d; x)\right|\right)_x 
	= \left| \frac{2\sqrt{2}}{w_0} \frac{\sgn(\zeta_d)}{\sqrt{1+1/\zeta_d^2}}\Delta x\right|.
\end{equation}
See the \appendixname~\ref{apx:derivationMaxPsXi} for the detailed derivation of this equation.
To aid intuitive understanding, $\Delta \phaseps (\zeta_d; x)$ at the PSF periphery (i.e., $x = w(z_d)/\sqrt{2}$) is plotted as a function of $\zeta_d$ in Fig.\@ \ref{fig:piMaxPhaseCurvDependence}.  

Based on this form of the equation, it is evident that this ``maximum value of the absolute phase increment'' reaches a maximum at $\zeta_d \rightarrow \pm \infty$, i.e., when $z_d \rightarrow \pm \infty$, and is given by
\begin{equation}
	\label{eq:maxMaxPhasePs}
	\mathrm{max}\left( \mathrm{max}\left(\left|\Delta \phaseps (\zeta_d; x)\right|\right)_x\right)_{\zeta_d}
	=\lim_{\zeta_d \rightarrow \pm \infty} \left| \frac{2\sqrt{2}}{w_0} \frac{\sgn(\zeta_d)}{\sqrt{1+1/\zeta_d^2}}\Delta x\right|
	= \frac{2\sqrt{2}}{w_0} \Delta x.
\end{equation}
Because we took the maximum of the maximum, the right hand side of this equation represents the maximum absolute phase increment that can occur when we sample the OCT signal with a sampling distance of $\Delta x$.

To fulfill the Nyquist condition and thus ensure that the sampled OCT signal can be refocused using holographic refocusing methods, the value of Eq.\@ (\ref{eq:maxMaxPhasePs}) should be smaller or equal to $\pi$.
This gives us the following criterion (i.e., the Nyquist criterion) for the holographic refocusing process.
\begin{equation}
	\Delta x \leq  \frac{\pi w_0}{2\sqrt{2}}.
\end{equation}
This criterion can be interpreted as follows.
Specifically, as long as the adjacent-pixel separation is smaller than $\pi / 2$ times the $1/e$-radius of the diffraction-limit PSF amplitude ($w_{0}/\sqrt{2}$), the defocus is correctable via holographic refocusing, regardless of the defocus distance.
In other words, as long as the adjacent-pixel separation is smaller than $\pi/4$ times the 
lateral resolution, i.e., $1/e^2$-diameter of the diffraction-limited PSF intensity ($\sqrt{2}w_{0}$), the defocus remains correctable.

\subsubsection{Nyquist criterion for \scFFOCT}
The Nyquist criterion for \scFFOCT can be derived by following the same logic used in the point-scanning OCT case, but starting with the phase-only function of Eq.\@ (\ref{eq:psfPhaseSc}) and the radius of the PSF of $x=w(z_d)$, which is the $1/e$-radius of the amplitude of Eq.\@ (\ref{eq:psfAmpSc}).

The absolute phase increment between adjacent pixels for \scFFOCT is given by
\begin{equation}
	\label{eq:ScPhaseDerivative}
	\left|\Delta \phasesc(x;z_d)\right|
	= \left| \frac{\partial}{\partial x}\left( \frac{nk_0 x^2}{2 R(z_d)} \right) \Delta x\right|
	= \left|\frac{nk_0 x}{R(z_d)} \Delta x\right|,
\end{equation}
and the maximum absolute phase increment at the normalized defocus depth of $\zeta_d$ is
\begin{equation}
	\label{eq:MaxScXi}
	\mathrm{max}\left(\left|\Delta \phasesc (\zeta_d; x)\right|\right)_x 
	= \left| \frac{2}{w_0} \frac{\sgn(\zeta_d)}{\sqrt{1+1/\zeta_d^2}}\Delta x\right|.
\end{equation}
The non-absolute version of this equation is plotted in Fig.\@ \ref{fig:piMaxPhaseCurvDependence} (orange curve).
Is it can be seen in the plot, the maximum absolute phase increment approaches its maxima asymptotically as $\zeta_d$ approaches $\pm \infty$ as
\begin{equation}
	\label{eq:maxMaxPhaseSc}
	\mathrm{max}\left( \mathrm{max}\left(\left|\Delta \phasesc (\zeta_d; x)\right|\right)_x\right)_{\zeta_d}
	=\lim_{\zeta_d \rightarrow \pm \infty} \left| \frac{2}{w_0} \frac{\sgn(\zeta_d)}{\sqrt{1+1/\zeta_d^2}}\Delta x\right|
	= \frac{2}{w_0} \Delta x.
\end{equation}
To fulfill the Nyquist condition, the value must be smaller than or equal to $\pi$, and this gives us the Nyquist criterion for holographic refocusing for \scFFOCT, as follows
\begin{equation}
	\Delta x \leq  \frac{\pi w_0}{2}.
\end{equation}
These Nyquist criteria suggest that, regardless of the defocus distance, the defocus is correctable as long as the adjacent pixel separation remains smaller than or equal to $\pi / 4$ times the $1/e^2$-width of the 
diffraction-limit PSF intensity ($2w_0$).

It should be noted here that \scFFOCT does not have a confocal pinhole and is thus free from the confocality limit that will be discussed in the next section.
Therefore, this Nyquist criterion is only the requirement to ensure that the defocus is correctable for \scFFOCT.

\subsection{Confocality-limit criterion for point-scanning OCT}
\label{sec:confocalLimit}
In addition to the lateral sampling density limit, the maximum correctable defocus for point-scanning OCT is also limited by defocus-dependent optical loss by a confocal pinhole.
In other words, a greater defocus causes a stronger optical loss and a lower SNR.
If the SNR becomes too low, the image will no longer be observed, even it has been sharpened via holographic refocusing.

We assume that the total signal intensity ($\ips$) captured at a specific depth in the point-scanning OCT is proportional to a confocal function $h(z_d)$. The confocal function is defined as an intensity integral of PSF for point-scanning OCT over the lateral integration direction, as follows
\begin{equation}
	\label{eq:confocality}
	\ips(z_d)\propto h(z_d) \propto \int_{0}^{2\pi} \int_{0}^{+\infty} \left|\psfps(r)\right|^2 r \,\mathrm{d}r\, \mathrm{d}\theta
	= \frac{{\pi w_0}^2}{4[1+(z_d/z_R)^2]},
\end{equation}
where $\psfps(r)$ is the PSF defined by using Eq.\@ (\ref{eq:psfPsBroken}) along with the substitution of $\rr\cdot\rr = r^2$.
Here, we did not take the light attenuation by scattering from the sample into account for simplicity.
The details of this issue will be discussed in Section \ref{sec:discussionConfocalFunction}.

\begin{figure}
	\centering\includegraphics[width=0.6 \textwidth]{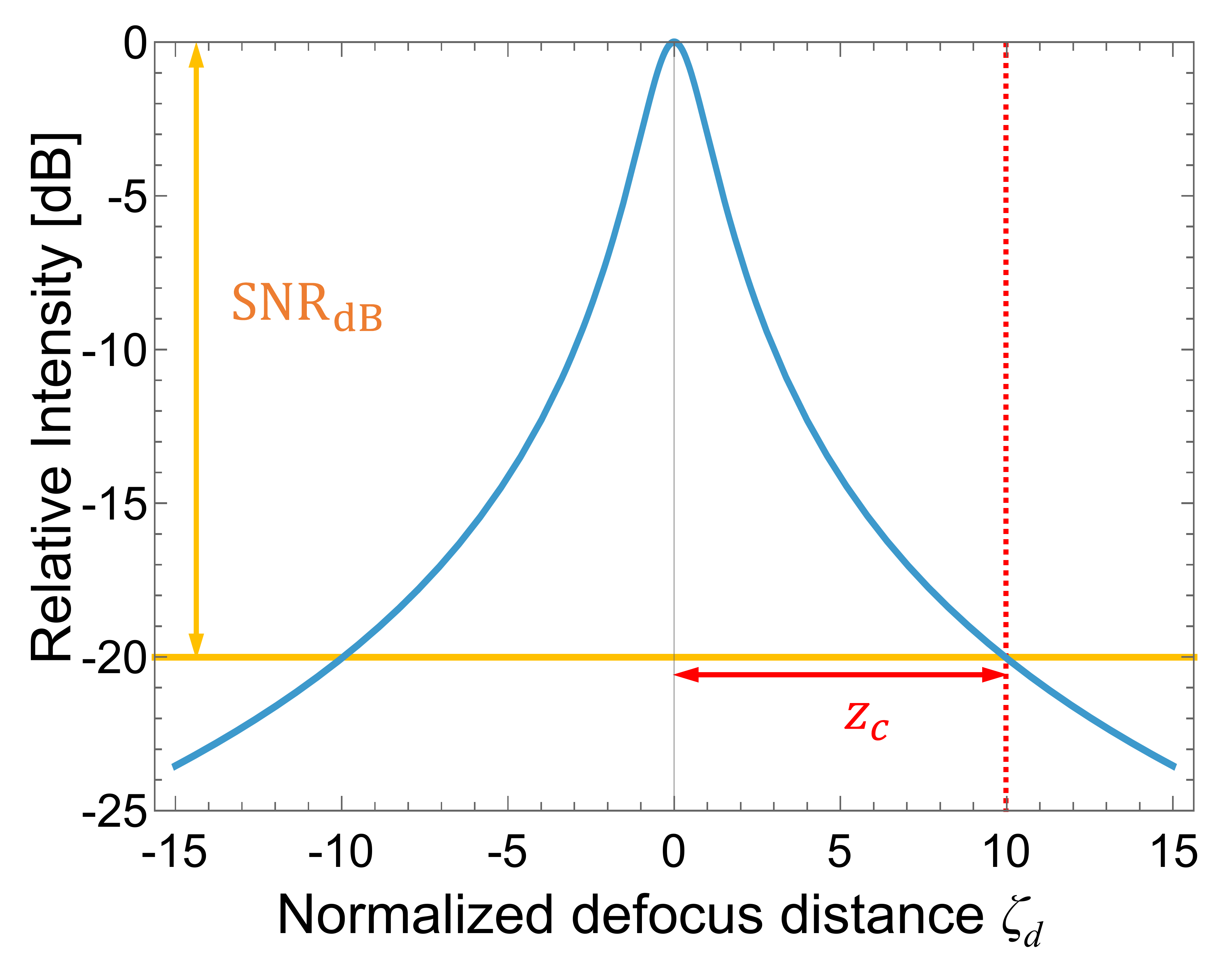}
	\caption{Intensity profile of the peak intensity of the refocused signal.
		The orange line and the red dashed line indicate the noise level and the critical defocus distance, respectively.
		The signal becomes observable after refocusing if the defocus distance is less than the critical defocus distance.
	}
	\label{fig:IntensityMCD}
\end{figure}
The dB-scaled profile of Eq.\@ (\ref{eq:confocality}) is shown as a function of the normalized defocus distance $\zeta_d = z_d/z_R$ in Fig.\@ \ref{fig:IntensityMCD}, where the peak at the in-focus depth is set as 0 dB.
This profile can be regarded as the peak intensity profile of refocused signal.
By assuming a specific SNR, we can then find the critical defocus distance $z_c$ at which the SNR becomes 0 dB and the signal disappears using
\begin{equation}
	10 \log_{10}\frac{1}{1+ \zeta_c^2} = - \mathrm{SNR_{dB}},
\end{equation}
where $\zeta_c = z_c/z_R$ is the normalized critical defocus and $\mathrm{SNR_{dB}}$ is the SNR in dB scale.
This definition of the critical defocus, i.e., the confocality-limit criterion, can be rewritten as
\begin{equation}
	\label{eq:criticalDefocus}
	z_c = z_R \sqrt{10^{\mathrm{SNR_{dB}}/10} -1 }.
\end{equation}
This definition of confocality-limit criterion is also illustrated schematically in Fig.\@ \ref{fig:IntensityMCD}.
Here, the dashed yellow line represents the noise level and the red dashed lines indicate the critical defocus distance.

By assuming that the sensitivity of the system is 100 dB and that sample attenuation is -60 dB, i.e., the SNR is 40 dB, the critical defocus distance $z_c$ then becomes approximately $100 z_R$.
Similarly, for an SNR of 20 dB, the critical defocus distance $z_c$ is approximately $10 z_R$.

\subsection{Summary for maximum sampling distance}
\label{sec:summaryMCD}
For point-scanning OCT, the maximum correctable defocus is limited by the more stringent of two criteria; the Nyquist criterion and the confocality-limit criterion.
According the Nyquist criterion, the defocus remains correctable as long as the pixel separation is smaller than $\pi/4$ times the diffraction limit of the 
lateral resolution defined as $1/e^2$-width of the PSF intensity
regardless of the defocus distance.
According to the confocality-limit criterion, the defocus distance should be smaller than the critical defocus distance $z_c$ defined in Eq.\@ (\ref{eq:criticalDefocus}), otherwise the signal cannot be observed, even after holographic refocusing.

For \scFFOCT, the Nyquist criterion is only one criterion.
The final description of this criterion becomes the same with that of the point-scanning OCT.
Namely, the defocus is correctable as long as the horizontal and vertical pixel separation is smaller than $\pi / 4$ times the 
lateral diffraction-limit resolution
.
Note that the Nyquist criteria for both types of OCT does not depend on the defocus distance.
Unlike point-scanning OCT, \scFFOCT does not employ confocal gating, meaning that out-of-focus light is still captured and can be computationally refocused.
The absence of the confocality make \scFFOCT advantageous for computational refocusing.

\section{Examples cases}
The maximum correctable defocus and the related system specifications have been analyzed for several OCT systems, with results as summarized in Table \ref{tab:systems}.
A Jones-matrix swept-source OCT (JM-SSOCT) system constructed by the authors \cite{Lida2022BOE} was included as a representative of scanning swept-source OCT with relatively low lateral resolution.
Although this system is polarization sensitive, it does not affect our analyses.
Please note that the 
lateral resolution was defined as $1/e^{2}$ of the beam spot diameter, or equivalently, as the $1/e^{2}$-width of the PSF amplitude, in Ref.\@ \cite{Lida2022BOE} and was 18 \um.
On the other hand, in the present manuscript, the 
lateral resolution is defined as the $1/e^{2}$-width of the PSF intensity 
($\sqrt{2}w_{0}$)
.
And hence, the 
lateral resolution of this JM-SSOCT is $18/\sqrt{2}$ \um (that is 12.7 \um) in the definition of the present manuscript.

A standard spectral-domain OCT (SD-OCT) system operating in the 840-nm band built by the authors \cite{Oida2021BOE,Morishita2023BOE} was included as a representative example of a relatively high-resolution scanning OCT system.
Although the lateral resolution of this system is written as 4.9 \um in Refs.\@ \cite{Oida2021BOE,Morishita2023BOE}, that must be $4.9/\sqrt{2}$ \um (that is 3.5 \um) in the definition of the present manuscript.

These two systems have been used for the studies of holographic signal processing and computational refocusing \cite{Lida2022BOE,Oikawa2020BOE,Tomita2023BOE}.

As an example of \scFFOCT system, we included our own \scFFOCT system\cite{Nobuhisa2025BiOS} because this system follows a standard \scFFOCT configuration and all system design parameters and specifications are available.
One important variation of \scFFOCT is off-axis \scFFOCT.
And hence, time-domain \cite{Sudkamp2016OL} and swept-source \cite{Hillmann2017OE} off-axis \scFFOCT are included in our analysis.

We also include spatio-temporal-optical-coherence tomography (STOC-T) demonstrated by Auksorius \etal \cite{Auksorius2019BOE,auksorius_vivo_2020}.
The STOC-T system has also been used for holographic aberration correction \cite{Borycki2020OL}.
Since STOC-T can be regarded as a \scFFOCT when the wavefront modulation is turned off, we included their optical parameters in the table for assessing the optical configuration.

In Table \ref{tab:systems}, the 
lateral resolutions of the point-scanning OCT systems were shown as the $1/e^2$-width of the PSF intensity 
($2w_0$)
.
In the case of \scFFOCT, the lateral resolution was calculated with the formula $4 \times 0.22 \frac{\lambda}{\mathrm{NA}_{\mathrm{col}}}$ according to the Gaussian profile fitting to the Airy disk pattern (Eq.~(21) in Ref.~\cite{Zhang2007AO}), where NA$_{\mathrm{col}}$ is the collection NA\@.

Note that the Nyquist criterion for both point-scanning OCT and \scFFOCT can be summarized as, ``if the ratio of pixel separation to lateral resolution (i.e., fractional pixel separation) is smaller than or equal to 78.5\% (i.e., $\pi/4$), the defocus can be corrected regardless of the original defocus amount.''
In the Jones-matrix SS-OCT with the a configuration of 512 $\times$ 512 lateral pixels with a 3 mm $\times$ 3 mm or smaller field of view, this criterion is fulfilled.
The scanning SD-OCT has higher lateral resolution than the Jones-matrix SS-OCT, and thus it requires smaller field of view, such as  1 mm $\times$ 1 mm, as long as far as we kept the same lateral pixel number; 512 $\times$ 512 pixels.
The on-axis SS-FFOCT of the University of Tsukuba was designed to sufficiently fulfill this criterion.
The STOC-T systems have a fractional pixel separation fulfilling the criteria.
The off-axis SS-FFOCT of the University of L\"ubeck also fulfills the criterion, but the criterion becomes tighter in practice for the off-axis configuration as we will discuss in Section \ref{sec:lateralPhaseModulation}.

\begin{table}[h]
	\caption{Specifications and critical defocus distances for example OCT systems.
		Pix., res., sep., and FOV are abbreviations for pixel, resolution, separation, and field of view, respectively.
		The critical defocus distance is the confocality-limit-based critical defocus distance that was defined in Section \ref{sec:confocalLimit}.
}
	\centering\includegraphics[width=13cm]{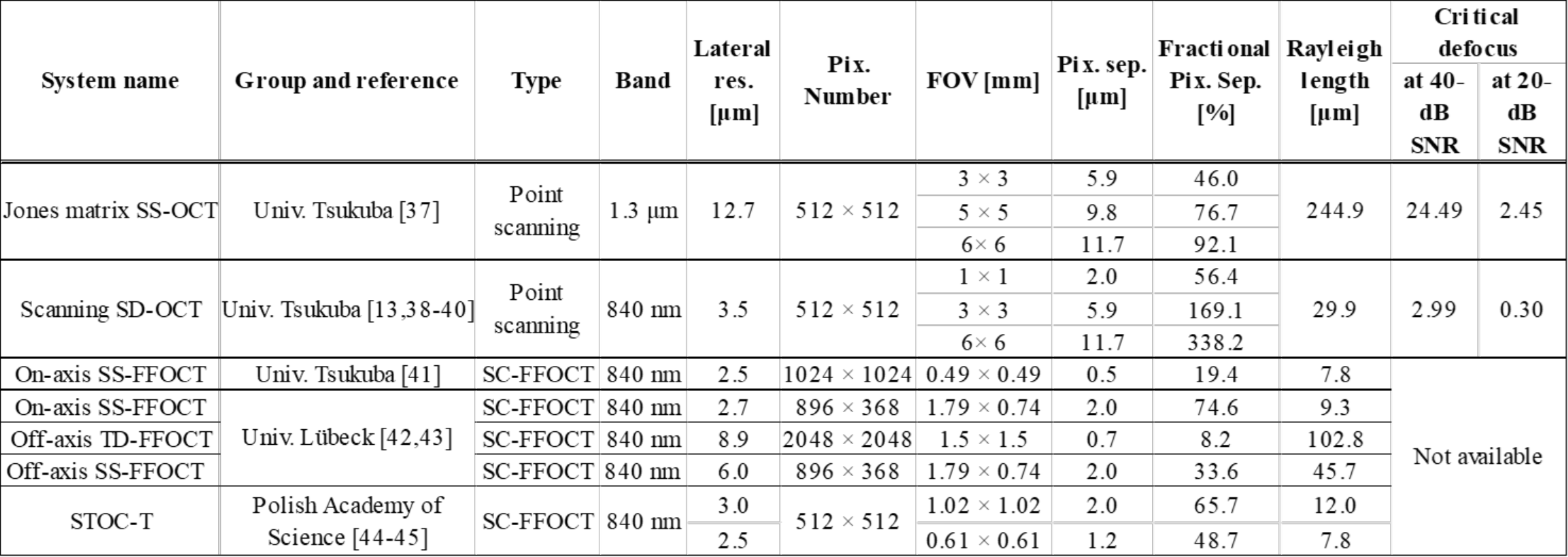}
	\label{tab:systems}
\end{table}

\section{Discussion and conclusions}

\subsection{Impact of lateral phase modulation by off-axis reference and BM-scan}
\label{sec:lateralPhaseModulation}
In some OCT systems, the OCT images are intentionally laterally modulated.
Among the existing \scFFOCT methods, the off-axis SS-FFOCT system of the University of L\"ubeck uses a tilted reference beam\cite{Hillmann2017OE}.
For point-scanning OCT, the required modulation can be achieved via simultaneous reference modulation with the transversal scan, e.g., BM-mode scan\cite{Yasuno2006AO}, and/or off-pivot use of a galvanometer scanner\cite{Hitzenberger2015JMO}.
In these cases, the modulation causes a spatial carrier frequency shift, and thus the spatial frequency spectrum of the OCT image is shifted into the high-frequency region. 
This may lead to stricter requirements for the lateral sampling density, and thus the Nyquist criterion may become tighter.

It might be important to analyze these effects theoretically in future work.

\subsection{Limitations and solutions}
The lack of confocality makes \scFFOCT method advantageous for computational refocusing, but it can also cause image degradation because of the multiple-scattering (MS) signals.
This problem can be resolved using methods that combine hardware modification with signal processing.
For example, STOC-T, which is a variation of \scFFOCT, overcomes the MS-signal related image degradation by using wavefront modulation and subsequent incoherent image averaging\cite{Tomczewski2022BOE}.
Multi-focus averaging (MFA) methods \cite{Lida2024BOE, Lida2023BOE, Yiqiang2024BOE} represent a combination of sequential OCT image acquisition with focus position modulation using an electrically tunable lens with subsequent computational refocusing and complex averaging. 
Although these methods were demonstrated with respect to point-scanning OCT, they can also be applied to \scFFOCT.

\label{sec:discussionConfocalFunction}
Another factor thats affect the imaging depth but is not considered in our analyses is signal attenuation caused by scattering and absorption characteristics of the sample.
Specifically, even the measurement fulfills the Nyquist criterion and the confocality-limit criterion, the signals cannot be observed if the signal attenuation is too high.

Although the optical property and the structure of the sample can affects the confocal function, this effect is not recapitulated by the na\"ive confocal function used in our analyses [Eq.\@ (\ref{eq:confocality})].
Some studies are dedicated to define and/or measure more accurate depth-intensity profile \cite{Smith2015IEEE, deBoer2021BOE}, and they can be used to enhance the accuracy of our analysis.

Another limitation in our analyses involves the usage of approximations in the pupil and spot descriptions.
For point-scanning OCT, the modeling was based on the paraxial Gaussian model \cite{Zhang2007AO}, which tacitly assumes that the lenses are aplanatic, and this is not an accurate approximation for very high NA cases.
For \scFFOCT, we approximated the collection spot using a Gaussian spot
(see Section \ref{sec:psfsc}).
In other words, we approximated the collection pupil using a Gaussian pupil, whereas it is a cylinder function with a clear cut-off frequency in reality.
These approximations are reasonable for most of the realistic cases, but some modification may be required to apply the analyses to very high NA cases.

In section \@\ref{sec:psfsc}, the illumination spot of the \scFFOCT is a constant, i.e., the uniform illumination.
It should be noted here that, although the illumination field (i.e., the illumination spot) does not extend infinitely largely, it can be reasonably considered as a constant as far as the illumination field is sufficiently large.

The \scFFOCT is advantageous when compared with point-scanning OCT in terms of the phase stability because of its parallel detection nature.
It may be worthwhile to analyze the effects of the phase stability, and those of the sample motion, on the computational refocusing performance in future work.

The present theory is limited to a two-dimensional lateral analysis.
The modification of the theory by incorporating spatial three-dimensional pupil-based imaging theory \cite{Villiger2010JOSA} or four-dimensional (i.e., space and time) imaging theory \cite{fukutake_four-dimensional_2025-1,fukutake_unified_2025-1} may improve the accuracy and applicability of our analyses.

\subsection{Spatially incoherent full-field OCT}
Our analysis of FF-OCT was limited to the spatially coherent cases only.
However, most time-domain FF-OCT uses spatially incoherent light.
It has been noted that this incoherent nature results in virtual pinhole effects\cite{Tricoli2019JOSA}, and thus spatially-incoherent FF-OCT may be affected by the confocality limit.
It may thus be important to extend our theoretical analyses to these spatially-incoherent cases in future work.

\subsection{Conclusion}
In this paper, a theoretical consideration of the limitations of holographic refocusing has been presented, and two types of criteria, i.e., the Nyquist criterion and confocality-limit criterion, were derived.
Specifically, point-scanning OCT and \scFFOCT methods were modeled using a dual pupil-based formulation to derive their Nyquist criterion.
The Nyquist criterion give the required sampling densities for holographic refocusing, and can be summarized as follows: ``the defocus is correctable regardless of the defocus amount as long as the lateral pixel density (i.e., sampling density) is lower than 78.5 \% (i.e., $\pi/4$) of the lateral resolution.''
Here the lateral resolution is defined as 1/$e^2$-width of PSF intensity.
It should be noted that, in some literature, the lateral resolution is defined as $e^2$-width of the beam spot, which is $\sqrt{2}$ times larger than the 1/$e^2$-width of PSF intensity.

Unlike \scFFOCT, the point-scanning OCT is also restricted by the confocality-limit criterion. 
In summary, if the SNR is 40 dB or 20 dB, the signal becomes unobservable at a defocus distance of 100 times or 10 times of the Rayleigh range, respectively, even after refocusing.

In conclusion, the correctable defocus amount of the point-scanning OCT is highly limited by the confocality in practice, not only by the Nyquist criteria.
On the other hand, for \scFFOCT, computational refocusing can work well even for very large defocus distances, as far as the lateral sampling density and the resolution of the system were designed to fulfill the Nyquist criterion.
This indicates that \scFFOCT is a particularly suitable technique for optical coherence microscopy.

\appendix
\makeatletter 
	\def\@seccntformat#1{\appendixname\ \csname the#1\endcsname:\quad}
\makeatother

\section{Derivation of Eq.~(\ref{eq:MaxPsXi})}\label{apx:derivationMaxPsXi}
Eq.\@ (\ref{eq:MaxPsXi}) was derived from  Eq.\@ (\ref{eq:PS_max}) as follows.

By substituting the phase curvature $R(z_d)$ [Eq.\@ (\ref{eq:phaseCurvature})], beam radius $w(z_d)$ [Eq.\@ (\ref{eq:beamradius})],the in-focus beam radius $w_{0}$, the refractive index $n$, and the wavenumber $k_{0}$, Eq.\@ (\ref{eq:PS_max}) becomes
\begin{equation}
	\mathrm{max}\left(\left|\Delta \phaseps (z_d; x)\right|\right)_x
	= \left|\frac{\frac{2\sqrt{2} n\pi}{\lambda_0} w_0\sqrt{1+(z_d/z_R)^2}}{z_d \left[ 1+(z_R/z_d)^2\right]} \Delta x \right|.
\end{equation}
We then multiply the denominator by Rayleigh length $z_{R}$ and divide it by Rayleigh length $z_{R}$ at the same time.
But the former $z_{R}$ is expressed as $z_{R}=\frac{n w_0^2\pi}{\lambda_0}$, while the latter is directly expressed as $z_{R}$.
Namely, the denominator is multiplied by $\left(n w_0^2 \pi/\lambda_0\right)/z_R$ as
\begin{equation}
	\mathrm{max}\left(\left|\Delta \phaseps (z_d; x)\right|\right)_x
	= \left|\frac{2\sqrt{2}}{w_0}\frac{z_R \sqrt{1+(z_d/z_R)^2}}{z_d\left[ 1+(z_R/z_d)^2\right]} \Delta x \right|.
\end{equation}
This equation can be rewritten as 
\begin{equation}
	\mathrm{max}\left(\left|\Delta \phaseps (z_d; x)\right|\right)_x
	= \left|\frac{2\sqrt{2}}{w_0} \frac{\sgn(z_d)}{\sqrt{1+1/(z_d/z_R)^2}}\Delta x\right|.
\end{equation}
Eq.\@ (\ref{eq:MaxPsXi}) is finally derived by introducing a normalized defocus distance, $\zeta_{d}=z_{d}/z_{R}$, into this equation.

\section*{Funding}
Core Research for Evolutional Science and Technology (JPMJCR2105);
Japan Society for the Promotion of Science (23KF0186, 21H01836, 22F22355, 22KF0058, 22K04962, 24KJ0510);
Chinese Scholarship Council (202106845011);
National Natural Science Foundation of China (62005123);
Natural Science Foundation of Jiangsu Province (BK20190455).

\section*{Disclosures}
Makita, Yasuno: Nikon (F), Nidek (F), Santec (F), Sky Technology(F), Panasonic (F), Topcon (F), Kao Corp.(F).
Fukutake: Nikon (E).

\section*{Data Availability}
Data underlying the results presented in this paper are not publicly available at this time but may be obtained from the authors upon reasonable request.

\bibliography{reference}

\end{document}